\documentclass[aps,11pt]{revtex4}
\usepackage{amsmath}
\usepackage{amsfonts}
\usepackage{amssymb}
\usepackage{graphicx}

\begin{document}
\title{Phase-space origin of superfluid stability in ring Bose-Einstein condensates}

\author{M. O. C. Pires}
\email{marcelo.pires@ufabc.edu.br}
\affiliation{Centro de Ci\^{e}ncias Naturais e Humanas, Universidade  Federal do ABC, Rua Santa Ad\'elia 166, 09210-170, Santo Andr\'{e}, SP,  Brazil}

\begin{abstract}

We present a kinetic description of superfluid currents in ring-shaped Bose-Einstein condensates based on the Wigner phase-space formalism. Starting from the Gross-Pitaevskii equation in a toroidal geometry, we derive a Vlasov-type equation for the angular Wigner function, in which the mean-field interaction generates an effective force proportional to the density gradient. Within this framework, we obtain the dispersion relation of collective modes and recover the Bogoliubov spectrum in the long-wavelength limit.

We show that the Landau criterion for superfluidity can be interpreted as the absence of resonant phase-space trajectories satisfying the condition \(\omega = q v_\ell\). In a ring geometry, the quantization of angular momentum leads to a discrete set of velocities, which suppresses the availability of resonant states and strongly inhibits Landau damping. In contrast, in the continuous limit \(R \to \infty\), the spectrum becomes quasi-continuous and the standard Landau damping mechanism is recovered, establishing a direct connection between kinetic resonances and the energetic criterion for superfluidity.

We further analyze the role of Bogoliubov depletion by considering a finite-width angular momentum distribution. Although resonant states formally exist in this case, we show that, for flow velocities below the sound velocity, the phase-space distribution does not provide the gradients required for energy transfer, and the superfluid current remains dynamically stable. Our results provide a unified phase-space interpretation of superfluidity, highlighting the role of angular momentum quantization and the structure of the distribution function in determining the stability of persistent currents.

\end{abstract}


\maketitle

\section{Introduction}

Persistent currents in superfluid systems represent one of the most striking manifestations of macroscopic quantum coherence. In ring-shaped Bose-Einstein condensates, the quantization of angular momentum gives rise to long-lived flow states that are remarkably robust against dissipation \cite{ram99,ryu07,she11,stu12}. 

The stability of these currents is commonly understood in terms of the Landau criterion \cite{lan80}, which states that superfluid flow is protected as long as the velocity remains below the critical value set by the speed of sound. In a toroidal geometry, this condition implies that a flow state can persist for long times as a quantized current, characterized by a winding of the macroscopic wave function around the ring. The metastability of such persistent currents was further analyzed by Bloch \cite{blo73}, who showed that the periodicity of the excitation spectrum prevents their decay under appropriate conditions. Thus, this criterion is typically formulated in energetic terms, by analyzing the spectrum of elementary excitations and determining whether processes that lower the energy of the system are available.

Persistent currents in ring Bose-Einstein condensates are known to remain stable even in the presence of Bogoliubov depletion and other excitations. While this robustness is typically understood in energetic terms, a clear kinetic interpretation of superfluid stability in such systems is still lacking.

While this energetic perspective has proven highly successful, it does not directly address the problem from a dynamical or kinetic viewpoint. In particular, it is natural to ask whether the stability of superfluid currents can be understood in terms of phase-space dynamics, in analogy with kinetic theories developed in plasma physics and classical many-body systems. In such approaches, dissipation and instability arise from resonant interactions between collective modes and the underlying distribution function, as exemplified by Landau damping \cite{lan46,lif81}. This naturally raises a fundamental question: can the Landau criterion for superfluidity be reinterpreted as a condition on the availability of resonant phase-space trajectories?

In this work, we address this question by developing a kinetic description of a Bose-Einstein condensate confined in a ring geometry within the Wigner phase-space formalism. Starting from the Gross–Pitaevskii equation, we derive a Vlasov-type equation for the angular Wigner function, in which the mean-field interaction generates an effective force proportional to the density gradient. This formulation enables a direct analysis of collective excitations and stability in phase space. In particular, we show that the Bogoliubov dispersion relation emerges naturally in the long-wavelength limit, and that the Landau criterion can be interpreted as the absence of resonant phase-space trajectories satisfying the condition
\(\omega = q v_\ell\).

A key aspect of our analysis is the role of angular momentum quantization in the ring geometry. In contrast to homogeneous systems with continuous momentum, the discrete spectrum of angular momentum gives rise to a discrete set of velocities, which strongly constrains the availability of resonant states. As a result, Landau damping is suppressed in finite systems. By considering the continuous limit, in which the radius of the ring becomes large and the spectrum approaches a quasi-continuum, we recover the standard Landau damping mechanism and establish a direct connection between kinetic resonances and the energetic criterion for superfluidity.

We further investigate the effect of Bogoliubov depletion by incorporating the finite-width angular momentum distribution generated by interactions. Although resonant states formally exist in this case, we show that, for flow velocities below the sound velocity, the phase-space distribution does not provide the gradients required for energy transfer, and the superfluid current remains dynamically stable. This result provides a kinetic explanation for the persistence of superfluid currents despite the presence of depletion.

Overall, our results offer a unified phase-space interpretation of superfluidity, highlighting the role of resonant dynamics, angular momentum quantization, and the structure of the distribution function in determining the stability of persistent currents. These findings may also provide a kinetic framework for understanding the experimentally observed robustness of persistent currents in ring-shaped condensates.

The paper is organized as follows. In Sec. II we introduce the Gross-Pitaevskii equation in a ring geometry and define the relevant stationary states. In Sec. III we construct the Wigner function and derive the corresponding kinetic equation. In Sec. IV we analyze the linear response of the system and obtain the dispersion relation of collective modes. In Sec. V we reinterpret the Landau criterion as a condition on phase-space resonances. In Sec. VI we discuss the role of angular momentum quantization and the suppression of Landau damping in finite systems. In Sec. VII we consider the continuous limit and recover the standard Landau damping mechanism. In Sec. VIII we analyze the effect of Bogoliubov depletion and its implications for the stability of superfluid currents. Finally, in Sec. IX we present our conclusions.

\section{Gross-Pitaevskii equation in a ring}

We consider a weakly interacting Bose-Einstein condensate confined to a one-dimensional ring of radius \(R\). In the mean-field regime, the system is described by the Gross-Pitaevskii (GP) equation for the macroscopic wave function \(\psi(\theta,t)\), where \(\theta \in [-\pi,\pi]\) is the angular coordinate along the ring. The dynamics is governed by
\begin{equation}
i\hbar\, \partial_t \psi(\theta,t)=-\frac{\hbar^2}{2mR^2}\,\partial^2_\theta \psi(\theta,t)
+
g\,|\psi(\theta,t)|^2 \psi(\theta,t),
\label{GP_ring}
\end{equation}
where \(m\) is the particle mass and \(g=4\pi \hbar^2 a/m\) is the strength of the short-range interaction, and $a$ is the scattering length. 

The periodic boundary condition \(\psi(\theta+2\pi,t)=\psi(\theta,t)\) reflects the topology of the ring and leads to the quantization of angular momentum. Stationary solutions of Eq.(\ref{GP_ring}) can be written as
\begin{equation}
\psi(\theta,t) = \sqrt{n_0}\,e^{i(\ell\theta - \mu t/\hbar)},
\end{equation}
where \(n_0\) is the uniform density, \(\mu\) is the chemical potential, and \(\ell \in \mathbb{Z}\) is the quantized angular momentum quantum number. Substituting into Eq.~(\ref{GP_ring}) yields
\begin{equation}
\mu = \frac{\hbar^2 \ell^2}{2mR^2} + g n_0.
\end{equation}

Each value of \(\ell\) corresponds to a persistent current circulating around the ring. The associated superfluid velocity is given by
\begin{equation}
v_\ell = \frac{\hbar \ell}{mR}.
\label{superfluid_velocity}
\end{equation}
In what follows, it is convenient to express the kinetic term in Eq.(\ref{GP_ring}) in terms of this velocity scale. The discreteness of \(\ell\) implies that the set of accessible velocities is also discrete, a feature that will play a central role in the kinetic description developed in this work.

We will consider small perturbations around a given stationary state with angular momentum \(\ell_0\). Writing
\begin{equation}
\psi(\theta,t) = e^{i\ell_0\theta}\,\phi(\theta,t),
\end{equation}
and substituting into Eq.(\ref{GP_ring}), one obtains an equivalent GP equation for \(\phi(\theta,t)\) in a frame co-moving with velocity \(v_{\ell_0}\). This transformation will be useful when analyzing fluctuations and collective excitations around a given superflow.

The formulation above provides the starting point for the phase-space description introduced in the next section, where we construct the Wigner function associated with \(\psi(\theta,t)\) and derive the corresponding kinetic equation governing its evolution.

\section{Wigner function and kinetic equation}

To develop a phase-space description of the condensate dynamics, we introduce the Wigner function associated with the macroscopic wave function \(\psi(\theta,t)\). In the ring geometry, the natural phase space is spanned by the angular coordinate \(\theta\) and the discrete angular momentum index \(\ell \in \mathbb{Z}\). The Wigner function is defined as
\begin{equation}
W(\ell,\theta,t)=\int_{-\pi}^{\pi}
d\sigma \,e^{i\ell\sigma}
\psi^*\left(\theta-\frac{\sigma}{2},t\right)
\psi\left(\theta+\frac{\sigma}{2},t\right).
\label{Wigner_def}
\end{equation}

The Wigner formalism provides a natural bridge between quantum dynamics and phase-space descriptions \cite{wig32}. This definition is the natural analogue of the usual Wigner transform adapted to a compact coordinate and satisfies the hermiticity condition $W(-\ell,\theta,t)=W^*(\ell,\theta,t)$, which follows directly from its definition. This discrete hermiticity property reflects the compact nature of the configuration space and differs from the standard Wigner function defined in continuous phase space. As a consequence, physical observables obtained from $W(\ell,\theta,t)$, such as the local density,
\begin{equation}
n(\theta,t) = \sum_{\ell} W(\ell,\theta,t),
\end{equation}
are real. Similarly, the momentum (or angular momentum) distribution is obtained by integrating over \(\theta\).

The evolution equation for \(W(\ell,\theta,t)\) can be derived by substituting the Gross-Pitaevskii equation (\ref{GP_ring}) into Eq. (\ref{Wigner_def}) and performing straightforward algebra. The kinetic term generates a drift in phase space, while the nonlinear interaction produces a term that couples different values of \(\ell\). 

In general, the interaction term generates a nonlocal contribution in \(\ell\), reflecting the underlying quantum nature of the dynamics. In the long-wavelength regime, however, this nonlocal structure can be systematically expanded in gradients, leading to a local Vlasov-type kinetic equation. This approximation is valid when the characteristic wave number satisfies \(q\xi \ll 1\), where \(\xi = \hbar/\sqrt{2mgn_0}\) is the healing length, ensuring that the density and phase vary smoothly along the ring. In this regime, the nonlocal structure can be systematically expanded in gradients, leading to a local Vlasov-type kinetic equation of the form
\begin{equation}
\partial_t W(\ell,\theta,t) + v_\ell\,\partial_\theta W(\ell,\theta,t)-(\partial_\theta V)\,\Delta_\ell W(\ell,\theta,t)
= 0,
\label{Vlasov}
\end{equation}
where 
the mean-field potential is given by
\begin{equation}
V(\theta,t) = g\,n(\theta,t) = g \sum_{\ell} W(\ell,\theta,t),
\end{equation}
and
 \(\Delta_\ell\) denotes a finite-difference operator acting on the discrete variable \(\ell\). For instance, a symmetric choice is
\begin{equation}
\Delta_\ell W(\ell,\theta,t)=\frac{W(\ell+1,\theta,t) - W(\ell-1,\theta,t)}{2}.
\end{equation}

Equation (\ref{Vlasov}) has the structure of a Vlasov equation, widely used in plasma physics and kinetic theory to describe collective dynamics in many-body systems \cite{lif81} and has a clear classical interpretation: the Wigner function evolves as a distribution in phase space under the action of a self-consistent force
\begin{equation}
F(\theta,t) = -\partial_\theta V(\theta,t),
\end{equation}
while particles with angular momentum \(\ell\) move with velocity \(v_\ell\). The nonlinearity of the Gross-Pitaevskii equation is thus encoded in the self-consistent coupling between the distribution function and the mean-field potential.

It is important to emphasize that Eq.(\ref{Vlasov}) is obtained as a semiclassical approximation of the full quantum dynamics. Higher-order gradient corrections would introduce additional terms involving higher derivatives in \(\theta\) and \(\ell\), analogous to quantum corrections in the standard Wigner formalism. However, for the purposes of analyzing collective excitations and stability in the long-wavelength regime, the Vlasov equation (\ref{Vlasov}) provides an accurate and physically transparent description.

This kinetic formulation establishes a direct connection between the dynamics of the condensate and classical phase-space evolution, allowing us to interpret superfluid stability in terms of trajectories and resonances in the (\(\ell,\theta\)) space.

\section{Linear response and collective modes}

We now analyze the collective excitations of the system within the kinetic framework introduced above. To this end, we consider small perturbations around a stationary state characterized by a uniform density and a well-defined angular momentum \(\ell_0\). The corresponding Wigner function is
\begin{equation}
W_0(\ell) = n_0\,\delta_{\ell,\ell_0},
\end{equation}
where \(n_0\) is the average density.

We introduce a small fluctuation \(\delta W(\ell,\theta,t)\) such that
\begin{equation}
W(\ell,\theta,t) = W_0(\ell) + \delta W(\ell,\theta,t),
\end{equation}
and linearize the kinetic equation (\ref{Vlasov}) to first order in \(\delta W\). The mean-field potential is similarly expanded as
\begin{equation}
V(\theta,t) = g n_0 + g\,\delta n(\theta,t),
\end{equation}
with
\begin{equation}
\delta n(\theta,t) = \sum_\ell \delta W(\ell,\theta,t).
\end{equation}

Substituting into Eq.(\ref{Vlasov}) and retaining only linear terms, we obtain
\begin{equation}
\partial_t \delta W + v_\ell\, \partial_\theta \delta W
- g\,(\partial_\theta \delta n)\,\Delta_\ell W_0 = 0.
\label{linear_eq}
\end{equation}

We now look for plane-wave solutions of the form
\begin{equation}
\delta W(\ell,\theta,t)=\delta W(\ell)\,e^{i(q\theta - \omega t)},
\end{equation}
which leads to
\begin{equation}
(-\omega + q v_\ell)\,\delta W(\ell)= q g\,\delta n\,\Delta_\ell W_0(\ell).
\label{deltaW_eq}
\end{equation}

For the equilibrium distribution \(W_0(\ell)=n_0\,\delta_{\ell,\ell_0}\), the finite-difference operator yields
\begin{equation}
\Delta_\ell W_0(\ell)=\frac{n_0}{2}\left[
\delta_{\ell,\ell_0+1}-\delta_{\ell,\ell_0-1}
\right].
\end{equation}

Equation (\ref{deltaW_eq}) then shows that only the neighboring modes \(\ell=\ell_0 \pm 1\) are directly coupled by the perturbation. Summing Eq.(\ref{deltaW_eq}) over \(\ell\), we obtain a closed equation for \(\delta n\), which leads to the dispersion relation
\begin{equation}
1-g q \sum_\ell
\frac{\Delta_\ell W_0(\ell)}{\omega - q v_\ell}
= 0.
\label{dispersion_discrete}
\end{equation}

Substituting the explicit form of \(\Delta_\ell W_0\), the sum reduces to two contributions, yielding
\begin{equation}
1-\frac{g q n_0}{2}
\left[
\frac{1}{\omega - q v_{\ell_0+1}}-\frac{1}{\omega - q v_{\ell_0-1}}
\right]
= 0.
\end{equation}

This expression encodes the discrete nature of the phase space and shows that the collective dynamics arises from coupling between neighboring angular momentum states.

In order to make contact with the usual hydrodynamic description, it is convenient to consider the limit in which the variation of \(W(\ell)\) with \(\ell\) becomes smooth. In this case, the finite-difference operator can be approximated by a derivative,
\begin{equation}
\Delta_\ell W_0 \approx \partial_\ell W_0,
\end{equation}
and Eq.(\ref{dispersion_discrete}) reduces to
\begin{equation}
1-g q \sum_\ell
\frac{\partial_\ell W_0(\ell)}{\omega - q v_\ell}
= 0.
\label{dispersion_cont}
\end{equation}

For a sharply peaked distribution around \(\ell_0\), this expression can be evaluated analytically and leads to
\begin{equation}
(\omega - q v_{\ell_0})^2 = c_s^2 q^2,
\end{equation}
where the sound velocity is given by
\begin{equation}
c_s^2 = \frac{g n_0}{m}.
\end{equation}

This result coincides with the Bogoliubov dispersion relation in the long-wavelength limit \cite{lan80}, demonstrating that the kinetic formulation reproduces the standard collective modes of the condensate. In the following section, we use this framework to reinterpret the Landau criterion in terms of phase-space resonances.

\section{Landau criterion as a phase-space resonance condition}

The kinetic formulation developed in the previous sections provides a natural framework to reinterpret the Landau criterion for superfluidity in terms of phase-space dynamics. In this approach, collective excitations are described by solutions of the dispersion relation, while the response of the system is governed by the structure of the distribution function \(W_0(\ell)\) and its coupling to perturbations through resonant denominators of the form \((\omega - q v_\ell)^{-1}\).

A key observation is that the linear response exhibits singular behavior when the resonance condition
\begin{equation}
\omega = q v_\ell
\label{resonance_condition}
\end{equation}
is satisfied. This condition identifies phase-space trajectories whose velocity matches the phase velocity of the perturbation. In analogy with kinetic theories of plasmas, such resonances allow for energy exchange between collective modes and the underlying distribution function, providing a mechanism for damping or instability.

Substituting the dispersion relation obtained in the previous section,
\begin{equation}
\omega = q v_{\ell_0} \pm c_s q,
\label{dispersion_relation}
\end{equation}
into the resonance condition (\ref{resonance_condition}), one finds that resonant states must satisfy
\begin{equation}
v_\ell = v_{\ell_0} \pm c_s.
\label{resonant_velocity}
\end{equation}

This result provides a direct connection with the Landau criterion. If the flow velocity satisfies \(v_{\ell_0} < c_s\), the condition (\ref{resonant_velocity}) cannot be fulfilled by states within the physically accessible region of the distribution, and no resonant phase-space trajectories are available. Conversely, if \(v_{\ell_0} > c_s\), solutions to Eq.(\ref{resonant_velocity}) exist, and resonant coupling becomes possible.

This leads to the following interpretation: The Landau criterion corresponds to the absence of resonant phase-space trajectories. In this picture, superfluidity arises not from the absence of excitations per se, but from the impossibility of satisfying the resonance condition required for energy transfer. The breakdown of superfluidity, in turn, is associated with the appearance of accessible resonant states that enable the redistribution of energy between the collective mode and the microscopic degrees of freedom.

It is important to emphasize that the presence of a formal solution to the resonance condition does not necessarily imply dissipation. The actual transfer of energy depends on the structure of the distribution function, as encoded in quantities such as \(\partial_\ell W_0\). In particular, the sign and magnitude of this gradient determine whether resonant states contribute to damping or amplification of the perturbation.

Within this framework, the Landau criterion emerges as a kinematic constraint on the accessibility of resonant phase-space trajectories, rather than as a purely energetic condition. This interpretation provides a unified connection between the kinetic description developed here and the traditional formulation of superfluidity, and sets the stage for analyzing how discreteness and finite-size effects modify the availability of resonant channels in the following sections.

\section{Discrete phase space: ring geometry}

A central feature of the ring geometry is the discreteness of the angular momentum variable \(\ell \in \mathbb{Z}\), which implies that the phase space of the system is intrinsically discrete along the momentum direction. This property has important consequences for the kinetic description developed above, particularly regarding the availability of resonant phase-space trajectories.

In the kinetic framework, resonant processes are governed by the condition Eq.~(\ref{resonance_condition})
which identifies those states that can exchange energy with a collective excitation of frequency \(\omega\) and wave number \(q\). In a system with continuous momentum, this condition can typically be satisfied for a broad range of parameters, as the velocity \(v\) can take arbitrary values. As a result, resonant states are generically present, and mechanisms such as Landau damping naturally arise.

In contrast, in the ring geometry the allowed velocities are quantized according to Eq.~(\ref{superfluid_velocity})
so that the set of accessible velocities forms a discrete lattice in phase space. The resonance condition therefore reduces to a discrete matching problem, which may or may not admit solutions depending on the values of \(\omega\), \(q\), and \(R\). In particular, for a given collective mode, it is possible that no value of \(\ell\) satisfies the resonance condition, in which case no resonant phase-space trajectory is available.

This observation has a direct impact on the stability of superfluid currents. For a flow characterized by angular momentum \(\ell_0\), the dispersion relation yields Eq.~(\ref{dispersion_relation}).
The corresponding resonant velocities would be Eq.~(\ref{resonant_velocity}).
However, because \(v_\ell\) takes only discrete values, these conditions may not be exactly satisfied. When no integer \(\ell\) fulfills this relation, the resonant denominators in the linear response remain finite, and the mechanism responsible for Landau damping is effectively suppressed.

From this perspective, the robustness of persistent currents in a ring can be understood as a direct consequence of the discrete structure of phase space. Even in situations where the continuous theory would predict the existence of resonant states, the discrete spectrum can prevent their realization. In other words, the compact geometry imposes a constraint on the accessibility of phase-space trajectories that protects the system against kinetic dissipation.

It is worth noting that this suppression is not absolute, but depends on the relation between the system parameters. In particular, as the radius \(R\) increases, the spacing between consecutive velocities,
\begin{equation}
\Delta v = \frac{\hbar}{mR},
\end{equation}
decreases, and the spectrum becomes progressively denser. In this regime, the discrete set of allowed velocities approaches a quasi-continuum, and the probability of satisfying the resonance condition increases. This observation naturally leads to the continuous limit, which will be analyzed in the next section.

The discrete phase-space structure thus provides a natural explanation for the stability of superfluid currents in finite systems, and highlights the role of geometry in determining the availability of dissipative channels.

\section{Continuous limit and emergence of Landau damping}

The discrete structure of phase space discussed in the previous section is a direct consequence of the compact geometry of the ring. As the radius \(R\) increases, however, the spacing between consecutive angular momentum states decreases,
so that the set of accessible velocities becomes increasingly dense. In the limit \(R \to \infty\), the spectrum approaches a continuum and the discrete phase space effectively reduces to a continuous one.

In this regime, the finite-difference operator \(\Delta_\ell\) introduced in Eq.(\ref{Vlasov}) can be approximated by a derivative, and the kinetic equation reduces to the standard Vlasov equation in continuous phase space. Similarly, the dispersion relation (\ref{dispersion_discrete}) becomes
\begin{equation}
1 - g q \int d\ell \frac{\partial_\ell W_0(\ell)}{\omega - q v_\ell} = 0,
\label{dispersion_continuous}
\end{equation}
where the sum over \(\ell\) has been replaced by an integral.

The integral in Eq.(\ref{dispersion_continuous}) exhibits a singularity when the resonance condition \(\omega = q v_\ell\) is satisfied. Following the standard prescription, this singularity is handled by deforming the integration contour in the complex plane, leading to the decomposition
\begin{equation}
\frac{1}{\omega - q v_\ell}\,\to\,\mathcal{P}\frac{1}{\omega - q v_\ell}-i\pi \delta(\omega - q v_\ell),
\end{equation}
where \(\mathcal{P}\) denotes the Cauchy principal value. The imaginary part arising from the delta function gives rise to a finite damping rate, which can be written as
\begin{equation}
\gamma \propto - \partial_\ell W_0(\ell_*),
\end{equation}
%
within the standard Landau prescription, where \(\ell_*\) is determined by the resonance condition \(v_{\ell_*} = \omega/q\).

This result corresponds to the well-known Landau damping mechanism, originally introduced in plasma physics \cite{lan46} and later extended to a variety of many-body systems, including Bose gases \cite{fed98}, and in which energy is transferred between collective modes and resonant particles in phase space. In contrast to the discrete case, where the resonance condition may not be satisfied, the continuous spectrum ensures that a resonant state always exists, provided that the dispersion relation allows it. Landau damping appears as the continuum limit of the discrete phase-space dynamics.

The sign of the damping rate is controlled by the slope of the distribution function at the resonant point. For distributions that decrease with increasing \(\ell\), the derivative \(\partial_\ell W_0\) is negative, leading to damping of the collective mode. Conversely, if the distribution has a positive slope in the resonant region, the system becomes unstable and the perturbation grows exponentially.

\begin{figure}[t]
\centering
\includegraphics[width=0.9\columnwidth]{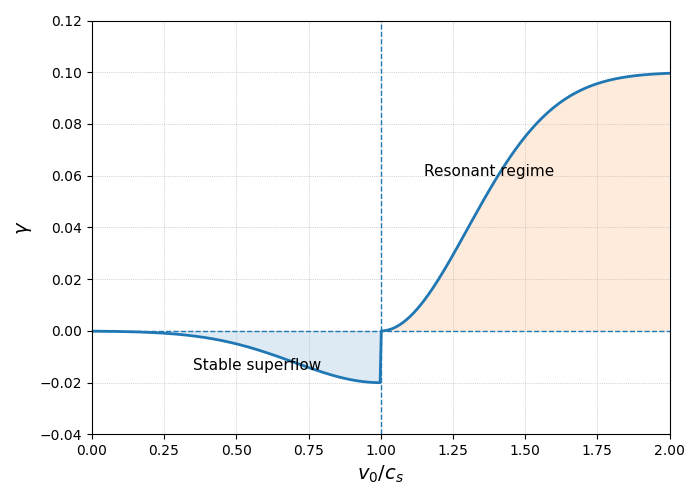}
\caption{
Stability diagram of persistent currents as a function of the normalized flow velocity $v_0/c_s$. For $v_0 < c_s$, no resonant phase-space trajectories are available and the flow remains stable. For $v_0 > c_s$, the resonance condition can be satisfied, leading to the onset of kinetic instability. This figure summarizes the connection between the Landau criterion and phase-space resonances.
The damping rate is computed using a smooth model distribution centered at $\ell_0$, capturing the qualitative effect of Bogoliubov depletion and illustrating how the emergence of resonant phase-space trajectories controls the stability of the flow.
}
\label{fig:stability}
\end{figure}

It is important to emphasize that the emergence of Landau damping in this limit is not due to a qualitative change in the underlying dynamics, but rather to the increased availability of resonant phase-space trajectories. The transition from a discrete to a continuous spectrum thus provides a natural explanation for the appearance of dissipative processes in large systems, and establishes a direct link between finite-size effects and kinetic stability. As illustrated in Fig.~\ref{fig:stability}, the onset of instability occurs when the flow velocity exceeds the sound velocity, marking the transition to a regime where resonant coupling becomes possible.

In the following section, we analyze how the structure of the equilibrium distribution, as determined by Bogoliubov theory, affects the efficiency of these resonant processes and the resulting stability of superfluid currents.

\section{Role of Bogoliubov depletion}

The analysis of the previous section shows that, in the continuous limit, resonant phase-space trajectories are generically available and give rise to Landau damping. However, the existence of resonant states alone is not sufficient to determine the stability of the system. The efficiency of energy transfer depends crucially on the structure of the underlying distribution function \(W_0(\ell)\), and in particular on its variation in the vicinity of the resonant point.

The role of the distribution function in determining damping has been extensively studied in the context of finite-temperature Bose gases \cite{zar99,pro08}. In a weakly interacting Bose-Einstein condensate, the equilibrium distribution is not strictly localized at a single angular momentum \(\ell_0\), but acquires a finite width due to quantum depletion. Within Bogoliubov theory, this corresponds to the occupation of excited modes with amplitudes determined by the coefficients \(v_k^2\) \cite{pro08}, leading to a distribution of the form
\begin{equation}
W_0(\ell) \simeq n_0,\delta_{\ell,\ell_0} + \delta W_{\text{dep}}(\ell),
\end{equation}
where \(\delta W_{\text{dep}}(\ell)\) describes the contribution of depleted particles around \(\ell_0\). This distribution is peaked at \(\ell_0\) and decreases away from it, with a characteristic width set by the interaction strength.

In the continuous limit, the resonance condition \(\omega = q v_\ell\) identifies a value \(\ell = \ell_*\) at which the collective mode can exchange energy with the distribution. In principle, the finite width of \(W_0(\ell)\) implies that such resonant states are always present. However, the contribution of these states to the damping rate is controlled by the gradient \(\partial_\ell W_0(\ell_*)\), which determines both the magnitude and the direction of energy transfer.

For flow velocities below the sound velocity, \(v_{\ell_0} < c_s\), the resonant point lies in a region where the distribution function is strongly suppressed. As a consequence, the slope \(\partial_\ell W_0(\ell_*)\) is small, and the corresponding contribution to the damping rate is negligible. More importantly, the structure of the distribution does not support a net transfer of energy from the collective mode to the excitations. Although resonant states formally exist, the phase-space distribution does not provide the gradients required for energy transfer.

This result shows that Bogoliubov depletion does not compromise the dynamical stability of superfluid currents. Instead, it leads to a situation in which the conditions for resonance are met in principle, but the corresponding processes are inefficient due to the small occupation of the relevant states. In this sense, superfluidity can be understood as arising not only from the absence of resonant trajectories, as in the discrete case, but also from the suppression of effective energy transfer in the continuous regime.

The analysis presented here thus provides a unified picture in which both finite-size effects and the structure of the equilibrium distribution contribute to the stability of superfluid flow. While discreteness suppresses the existence of resonant channels, the Bogoliubov distribution suppresses their effectiveness, ensuring that superfluid currents remain robust even when the spectrum becomes quasi-continuous.

\section{Unified interpretation}

The results obtained in the previous sections can be summarized within a unified phase-space framework that provides a coherent interpretation of superfluid stability across different regimes. In this picture, the dynamics of the condensate is described in terms of trajectories in phase space, and the stability of superfluid currents is determined by the availability and effectiveness of resonant channels that enable energy exchange between collective modes and the underlying distribution.

Three distinct regimes can be identified. In finite systems with discrete angular momentum, the phase space is intrinsically discrete and the resonance condition \(\omega = q v_\ell\) may not admit solutions. In this case, resonant phase-space trajectories are simply absent, and the system is protected against kinetic dissipation. This provides a natural explanation for the robustness of persistent currents in ring geometries.

In the opposite limit of large systems, the angular momentum spectrum becomes quasi-continuous and resonant states are generically available. In this regime, Landau damping emerges as a natural consequence of the kinetic description, with energy transfer governed by the structure of the distribution function near the resonant point. The stability of the system is then controlled by the sign and magnitude of the phase-space gradients, which determine whether perturbations are damped or amplified.

An intermediate situation arises when the distribution function has a finite width, as in the case of Bogoliubov depletion. Here, resonant states formally exist even in the absence of a strictly continuous spectrum. However, the efficiency of the corresponding processes is strongly reduced because the relevant regions of phase space are weakly populated. As a result, energy transfer remains suppressed and superfluid currents retain their stability.

These three regimes can be understood within a single conceptual framework in which superfluidity is associated with the effective absence of dissipative phase-space channels. This absence may arise either from the discrete nature of the spectrum, which prevents the existence of resonant trajectories, or from the structure of the distribution function, which suppresses the efficiency of energy exchange even when such trajectories are present.

From this perspective, the traditional energetic formulation of the Landau criterion and the kinetic description developed here appear as complementary views of the same underlying physics. The energetic condition identifies whether excitations can be created, while the kinetic approach determines whether these excitations can be accessed dynamically through resonant phase-space processes. The agreement between these viewpoints highlights the consistency of the phase-space interpretation and its ability to capture the essential features of superfluid behavior.

Finally, this framework suggests that similar ideas may be applicable to more complex systems in which the structure of phase space and the accessibility of trajectories play a central role. In particular, extending this approach to multicomponent systems may provide insight into the stability of coupled superflows, where the interplay between different distributions could lead to new types of resonant behavior. Such extensions represent a promising direction for future work.

\section{Conclusion}

In this work, we have developed a kinetic description of superfluid currents in ring-shaped Bose-Einstein condensates based on the Wigner phase-space formalism. Starting from the Gross-Pitaevskii equation, we derived a Vlasov-type equation for the angular Wigner function and showed that the dynamics of the condensate can be interpreted in terms of classical trajectories in phase space subject to a self-consistent mean-field force. Within this framework, the dispersion relation of collective excitations emerges naturally, recovering the Bogoliubov spectrum in the long-wavelength limit.

A central result of our analysis is the reinterpretation of the Landau criterion for superfluidity as the absence of resonant phase-space trajectories. In this picture, dissipation and instability arise from resonant conditions of the form \(\omega = q v_\ell\), which enable energy transfer between collective modes and the underlying distribution. In a ring geometry, however, the quantization of angular momentum leads to a discrete set of velocities, which strongly restricts the availability of such resonant channels. As a consequence, Landau damping is suppressed in finite systems, providing a kinetic explanation for the robustness of persistent currents.

By considering the continuous limit, in which the radius of the ring becomes large and the angular momentum spectrum approaches a quasi-continuum, we recovered the standard Landau damping mechanism. This establishes a direct connection between the kinetic phase-space description and the traditional energetic formulation of superfluidity. Furthermore, by incorporating the finite-width distribution associated with Bogoliubov depletion, we showed that although resonant states formally exist, the structure of the distribution does not support net energy transfer for flow velocities below the sound velocity. This demonstrates that the presence of depletion does not compromise the dynamical stability of superfluid currents.

Taken together, our results provide a unified phase-space interpretation of superfluidity, in which stability is determined by the structure of accessible trajectories in phase space and by the possibility of resonant energy exchange. In this view, superfluid behavior can be understood as the effective absence of dissipative channels, either due to spectral discreteness or to the properties of the underlying distribution function.

An interesting perspective opened by this approach is its potential extension to other systems where the structure of the spectrum and the accessibility of phase-space trajectories play a central role. In particular, it would be natural to explore analogous kinetic descriptions in lattice systems, where the competition between tunneling and interactions leads to qualitatively different regimes, such as the superfluid and Mott insulating phases. In such contexts, the transition between these phases may be interpreted in terms of changes in the connectivity and accessibility of phase-space trajectories, rather than solely through energetic considerations. Investigating whether a resonance-based kinetic criterion can be formulated for these systems, and how it relates to the breakdown of superfluid transport, represents an interesting direction for future work.

\end{document}